\documentclass[aps,prl,twocolumn,a4paper,showpacs]{revtex4}
\usepackage{amssymb,amsmath,graphicx,epsfig,rotate,epsf}

\begin{document}
\bibliographystyle{prsty}

\title{Dynamics of Turing patterns under spatio-temporal forcing}

\author{S. R\"udiger$^{1,2}$, D. G. M\'\i guez$^3$,
A. P. Mu\~nuzuri$^3$, F. Sagu\'es$^4$ and J. Casademunt$^1$}
 \affiliation{
$^1$Dept. E.C.M., Facultat de F\'{\i}sica, Universitat de Barcelona, Av.
Diagonal 647, 08028 Barcelona, Spain}

\affiliation{$^2$School of Computational Science and Information Technology,
F.S.U, Tallahassee Fl 32306, USA}

\affiliation {$^3$Facultade de F\'\i sica, Universidade de Santiago de
Compostela, 15782 Santiago de Compostela, Spain}

\affiliation{$^4$Departament de Qu\'{\i}mica F\'{\i}sica, Universitat de
Barcelona,
 Mart\'{\i} i Franqu\`es 1, 08028 Barcelona, Spain}

\begin{abstract}
We study, both theoretically and experimentally, the dynamical response of
Turing patterns to a spatio-temporal forcing in the form of a travelling wave
modulation of a control parameter. We show that from strictly spatial
resonance, it is possible to induce new, generic dynamical behaviors, including
temporally-modulated travelling waves and localized travelling soliton-like
solutions. The latter make contact with the soliton solutions of P. Coullet
Phys. Rev. Lett. {\bf 56}, 724 (1986) and provide a general framework which
includes them. The stability diagram for the different propagating modes in the
Lengyel-Epstein model is determined numerically. Direct observations of the
predicted solutions in experiments carried out with light modulations in the
photosensitive CDIMA reaction are also reported.

\end{abstract}
 \maketitle

        The study of pattern dynamics under external forcing provides
a powerful tool to deeply probe their inherently nonlinear mechanisms under
non-equilibrium conditions. A great deal of attention has been focused on
resonances or locking of spatially structured states, either stationary or
oscillatory, under temporal (spatially uniform) \cite {1,2,3,4,5} or spatial
(steady) modulations \cite {6,7,8,9,10}. Steady patterns in reaction-diffusion
systems typically arise from the celebrated Turing mechanism \cite {11,12}.
According to it, inhomogeneous distributions of chemical concentrations
self-organize spontaneously out of a non-structured medium as a result of a
competition between autocatalytic reaction steps and the differential
diffusivities of an activator (smaller) and an inhibitor-like (larger
diffusion) species. Turing patterns are endowed with an intrinsic wavelength,
depending only on the kinetic and diffusion parameters, but lack an intrinsic
frequency, in contrast to oscillatory chemical systems \cite {12}. Genuine
Turing patterns were first observed in quasi two-dimensional gel reactors
(pre-loaded with appropriate chemical indicators) in the CIMA \cite {13} and
the CDIMA \cite{16} reactions, and appear as patterned distributions of iodide.
The CDIMA reaction has the interesting feature of being photosensitive
\cite{14}.

It seems thus timely to search for generic behavior in the
unexplored perspective of spatio-temporal forcing of pattern
forming systems.
Specifically, we aim at studying the dynamical response of photosensitive
Turing patterns to the simplest external spatio-temporal forcing consisting of
a travelling-wave modulation of the control parameter associated to the
illumination. Through the mechanism of pure spatial resonance an external
frequency will thus be imposed in an otherwise nonoscillatory system. As a
consequence, new nontrivial dynamical modes are expected to arise which allow
to connect the two trivial limiting cases, namely: a travelling pattern locked
to the forcing at low velocities and a standing pattern resulting from the
time-averaging of the forcing  at large velocities.
 Analytical and numerical results will be reported
featuring the simplest of these spatio-temporal behaviors. Experiments
conducted with the CDIMA reaction are also provided which fully confirm our
theoretical predictions. Beyond the particular chemical context that motivates
our study, such solutions, and the conditions of their appearance, are
sufficiently generic to be applicable to a rather general class of pattern
forming systems including for instance Rayleigh-B\'enard convection. In a
sense, our study may be viewed as a development of the work of Coullet
\cite{7,15} on commensurate-incommensurate transitions in nonequilibrium
(spatially) forced systems.

The experimental system under study is modelled within the Lengyel-Epstein
scheme \cite{16}, once modified to include the effect of illumination, as
\cite{14}:
\begin{eqnarray}
\partial_t u&=& a -c u -4 \frac{u v}{1+u^2} - \phi + \frac{\partial^2 u}{ \partial x^2} \label{le1},\\
\partial_t v&=& \sigma(c u - \frac{u v}{1+u^2} + \phi + d \frac{\partial^2 v}{\partial x^2}). \label{le2}
\end{eqnarray}
Here $u$ and $v$ are the dimensionless concentrations of two of
the chemical species; $a$, $c$, $\sigma$, and $d$ denote
dimensionless parameters of the chemical system. The effect of
external illumination is introduced through the $\phi$-terms. This
contribution can be decomposed into the mean value $\phi_0$ and a
modulation part: $\phi(x,t)=\phi_0 + \epsilon \; \cos (k_f x + \omega t)$.
 For purely homogeneous illumination,
$\epsilon=0$, the equations admit a solution which in the
following will be referred to as base state: $u_0=(a-5 \phi_0)/(5
c)$, $v_0=a(1+u_0^2)/(5 u_0)$.

All our numerical results are obtained through integration of the
model reaction-diffusion equations (\ref{le1}),(\ref{le2}) with
periodic boundary conditions by means of a pseudo-spectral method
with a linear-implicit time-stepping. From here on we fix the
parameters to the following: $a=16, \; c=0.6, \; d=1.07, \;
\mbox{and} \; \sigma=301$. These values were chosen to reproduce
the experimental conditions referred to below. The large $\sigma$
value, corresponding to a strong diffusion contrast between the
two species, guarantees that we are far from the oscillatory
regime of this chemical system. Consequently, only the Turing
bifurcation will play a role. The remaining parameters concern
the forcing term. One of them, $\phi_0$, will serve as the
parameter to locate the position of the Turing bifurcation in the
homogeneous problem. For the given parameters it occurs at $\phi_0
\approx 2.3$ (the base state being unstable to Turing patterns
below $\phi_0 \approx 2.3$) and the critical wavenumber is $k_c
\approx 1.07$.

We begin our analysis with the case of exact 1:1 spatial resonance $k_c=k_f$.
Choosing the length $L$ of the periodic domain to be $10 \lambda$, where
$\lambda=2 \pi/k_c$ is the critical wavelength at instability, we fix $L=58.72$
and the amplitude $\epsilon=0.1$.

As is known from the analysis of time-independent forcing, the spatial 1:1
resonance yields an imperfect bifurcation to Turing patterns \cite{7}.
Accordingly, the base state ceases to be a stationary solution and is modified
into a non-homogeneous state for every value of the bifurcation parameter
$\phi_0$. For $\phi_0$ in the stable region (above 2.3) this state is a
travelling wave (TW). The TW locks to the forcing wave, adopting the same
wave-number and frequency, with only a constant phase difference. This is the
trivial state to be expected for slow driving. For large $\omega$ the amplitude
of the TW approaches 0.
The TW's exist to the right of the solid curve in fig.~\ref{diag1}.

Crossing the solid curve in fig.~\ref{diag1} the TW state
undergoes a first instability into a state with temporally
modulated amplitude (MTW, triangles in fig.~\ref{diag1}). This is
the signature of a Hopf bifurcation which introduces a new
frequency $f_H$ (see fig.~\ref{sim}a). Note that the modulation
of the amplitude occurs uniformly in the entire system. We found
that for large $\omega$ the Hopf frequency converges to the
frequency $\omega$ of the forcing wave.

The solutions and the transition described above can be rationalized in terms
 of an amplitude equation. Following standard envelope techniques near
 threshold \cite{7,Cross}, and with the forcing being
 invariant under the transformation $t \rightarrow t+T$, $x
\rightarrow x -\omega T/k_f$, for sufficiently small $\epsilon$ and $\omega$,
the slowly varying modulations of the travelling mode, in the case of perfect
1:1 resonance ($k_f=k_c$) will be given by the amplitude equation
\begin{equation}
\dot{A}= \mu A -|A|^2 A +\epsilon \exp{(-i \omega t)} +\partial ^2 A/\partial x^2.
\label{amp}
\end{equation}

Using polar coordinates, $A=R \exp{i \Theta}$, we look for homogeneous
solutions with $\Theta=\Theta_0 -\omega t$. As in the steady case for $\epsilon
\ne 0$ there is a non-zero solution for every $\mu$, the dimensionless distance
to threshold. Its amplitude approaches $0$ with increasing $\omega$ in
accordance with our observations for the Lengyel-Epstein model. $\Theta_0$ is
the phase shift between the forcing wave and the resulting pattern.

We further determine the stability of this solution with respect to homogeneous
perturbations. Directly from the amplitude equation:
\begin{eqnarray}
\dot{R}&=& \mu R -R^3 +\epsilon \; \mbox{cos}(\psi+\Theta_0),  \\
\dot{\psi}&=&-\frac{\epsilon}{R} \; \mbox{sin}(\psi+\Theta_0) + \omega,
\label{phase}
\end{eqnarray}
where we have defined: $\psi=\Theta+\omega t -\Theta_0$. Linearization about
the locked solution ($\psi=0$, $R=Q=const$) yields the following eigenvalues:
$\lambda=\mu-2 Q^2  \pm \sqrt{} (Q^4-\omega ^2)$. Corresponding to
fig.~\ref{diag1}, for large $\omega$, the marginal curve approaches the line
$\mu=0$, and the imaginary part of the eigenvalue, converges to the driving
frequency, as also observed numerically above~\cite{footnote2}.

We now address the more generic case of inexact 1:1 resonance~\cite{footnote3},
introducing a slight wavenumber misfit, $k_f \neq k_c$. To allow for a
continuous variation of the misfit in a finite system, we will fix the integer
wavenumber ($n=kL/2\pi$) of the forcing to $n=10$ and change smoothly the
length $L$ of the periodic domain. For example, for $L=65$ the 10th wave number
corresponds to $k_f \approx 0.97$. Fig.~\ref{mis} depicts the complex behavior
that was found changing $k_f$ from 0.9 to 1.26 ($L=70, \dots, 50$). The average
illumination $\phi_0$ was fixed to 2.25 during all of the simulations. For the
purely homogeneous forcing (i.e. $\phi^\prime=0$) this value corresponds to a
slightly unstable base state.

In fig.~\ref{mis} the stability domain in $\omega$ and $k_f$ is
given for four different states. The TW state, the solution locked
to the forcing, is the only stable solution for approximately
$\omega \le 1$. Above a roughly horizontal transition curve close
to $\omega=1.1$ the TW states are unstable. For $k_f=k_c$ and
close to this point
the transition occurs at $\omega \approx 1.15$
(cf. fig.~\ref{diag1}) and results in the MTW.

The domain of stability of the MTW is given by the solid curve in
fig.~\ref{mis}. Outside of this domain the attractors are either the TW state
or one of two different new localized states which we call soliton waves
(SOW)\cite{footnote1}.
A typical space-time plot is shown in fig.~\ref{sim}b. Spatial plots of these
states show that they resemble the soliton-like solutions for the case of
non-travelling forcing \cite{7} but they exhibit the following new properties:
the soliton, i.e., the localized suppression of the amplitude moves relatively
to the underlying travelling pattern with a relative velocity which may be
positive or negative. Furthermore, as the soliton travels along the pattern the
phase of the background pattern moves in either direction. In our case the
integer wavenumber of the background pattern is either $n=9$ or $n=11$ and the
corresponding states are represented in fig.~\ref{mis} with circles and
diamonds, respectively.

Beyond the range of forcing wavenumbers shown in fig.~\ref{mis}
there exist further soliton states in accordance with the
prediction for non-travelling forcing \cite{7}. Here we
will describe only the two states that are adjacent to the MTW
state in fig.~\ref{mis}. The states represented by diamonds
(resp. circles) carry a soliton that moves to the right (resp.
left).
The approximate range of stability
for these states is given by the dashed curve
(resp. dotted curve).

Note that the stability domains of
the soliton states and the MTW state overlap. Remarkably, at
sufficiently large $\omega$ the soliton states are even stable for
$k_f=k_c$, i.e., exact 1:1 resonance. This contrasts to the case
of static forcing \cite{7} where only the presence of a misfit can lead to
solitons.

The above localized states (SOW) can be described as quasi-periodic in the
sense that their behavior can be reproduced as superposition of two modes,
namely, the forcing (travelling) mode with wavelength $L/10$ and an adjacent
mode with wavelength $L/9$ or $L/11$, typically the closer one to the
characteristic wavelength of the Turing instability. The excited Turing mode is
standing so that the superposition of both yields a localized envelope
(soliton) moving to the right ($L/11$) or to the left ($L/9$). Furthermore, one
finds the velocity of the soliton to be proportional to the quotient of $\omega$ and
the difference of the wavenumbers of the two modes, which agrees with numerical
results of the full problem.

Note that the phase diagram depicted in fig.~\ref{mis} exhibits large regions
of multistability, with the corresponding hysteresis. It is worth remarking
that the stability boundaries here reported may well be affected by to
finite-size effects, since possible long wave-length, phase instabilities may
be prevented by the finite size of our simulations. A full envelope and
phase-diffusion description of this scenario in an infinite system deserves a
detailed study and will be addressed elsewhere.

A link between the SOW's and the solitons for steady forcing is apparent for
smaller $\epsilon$. We decreased the value of $\epsilon$ to 0.003 and
determined the domain of stability for the soliton 11 state for $\omega$ values
close to 0 (fig.~\ref{ton}). It appears that even for vanishing $\omega$ the
soliton states persist and the domain of stability forms a ``tongue'' with a
finite range at $\omega=0$. For larger $\epsilon$ and $\omega=0$ the SOW's do
not exist since the driving term dominates the equations and forces a
homogeneous pattern. Therefore it seems that the relative suppression of the
forcing term by a effective time-averaging due to a fast travelling forcing may
account for the existence of soliton states for larger $\epsilon$.

To prove the feasibility of controlled spatio-temporal forcing and to check the
validity and robustness of our predictions in a real system undergoing a Turing
instability, we have carried out experiments on the (photosensitive) CDIMA
chemical reaction, using the projection of travelling light patterns as the
controlled forcing. Experiments were performed in a thermostatized continuously
fed unstirred one-feeding-chamber reactor at 4 $\pm$ 0.5 ºC. Structures appear
in an agarose gel layer (2\% agarose, thickness 0.3 mm, diameter 20 mm). The
gel layer was separated from the feeding chamber by an Anapore membrane
(Whatman, pore size 0.2 mm) and a nitrocellulose membrane (Schleicher \&
Schnell, pore size 0.45 mm). Under the chosen set of reagent concentrations
(see caption of fig.5), the system at dark spontaneously yields disordered
stripe patterns with a wavelength of 0.54 mm. In a typical experiment, parallel
light stripes with a characteristic wavelength of precisely 0.54 mm were
focused on the gel layer and were moved in the horizontal direction with
well-controlled and constant velocity $v$. For very small values, trivial
travelling stripes, following adiabatically the imposed pattern, were found, as
expected. As the passing velocity was increased they readily transformed into
the modulated striped mode, as predicted by the theory (see fig.~\ref{exp}a).
Furthermore, when considering a slight misfit (see caption of fig.5), a
localized structure was observed, propagating in the opposite direction to that
of the stripes (see fig.~\ref{exp}b).This again constitutes a neat confirmation
of our theoretical findings. Further experiments are presently being conducted
to look for other modes of dynamical responses to such a spatio- temporal
modulation and will be published elsewhere.

We have described a generic mechanism to induce new pattern dynamics through
spatial resonance. The phenomenon is claimed to be generic for systems
undergoing a Turing instability. For instance we have already observed the same
type of response in the 1d Swift-Hohenberg equation. Within the framework of
chemical Turing patterns, the consideration of the 2d case is readily
accessible and points out to even richer phenomena.

S.R. is supported by the NSF under contract DMR-0100903. Financial support from
DGI (Spain) under projects BXX2000-0638-C02 and BFM2000-0348, and also from
European Commission under network HPRN-CT-2002-00312 is acknowledged.

\newpage
\begin{figure}[]
\caption{Type of attractor for
different values of $\omega$ and average illumination $\phi_0$:
boxes correspond to TW solutions, triangles to MTW solutions (see
text). The vertical line shows the position of the instability for
homogeneous forcing.\label{diag1}}
\end{figure}

\begin{figure}[]
\caption{Space-time plots of the
$u$-component of the modulated travelling wave solution for
$\omega=0.5$ and $\phi_0=2.11$ (a); and for the soliton solution
for $\omega=5$, $\phi_0=2.25$ and $k_f=0.9$ (b).} \label{sim}
\end{figure}

\begin{figure}[]
\caption{Phase
diagram in the space of $\omega$ (vertical) and $k_f$
(horizontal), filled box - TW, filled triangle - MTW  (see text
for notation), diamond - soliton state with wave length $L/11$,
circle - soliton state with wave length $L/9$. We have chosen
$\phi_0=2.25$. The value $k_f=k_c$ ($L=58.72$) corresponds to the
dashed vertical line showing the position of the perfect 1:1
resonance.} \label{mis}
\end{figure}

\begin{figure}[]
\caption{'Tongue' of soliton states as a function of $\omega$ for
small $\epsilon$. The dashed line approximates the boundary of the
soliton 'tongue' with $n=11$ as it approaches the $\omega=0$ line
(spatial forcing of Ref.\cite{7}). Here we have used
$\epsilon=0.003$.} \label{ton}
\end{figure}

\begin{figure}[]

\caption{ Experimental space-time
plots for the modulated travelling wave solution (a) and for the
soliton solution (b). The dashed line in (b) is a guide to the
eye. The input concentrations of reagents are 0.45 mM I2, 0.078 mM
ClO2, 10 mM H2SO4, 1.2 mM malonic acid and 10 mM polyvinil alcohol
with a residence time in the reactor of 250 s. With these
parameters the system spontaneously yields stripe patterns with a
wavelength of 0.54 mm. (Experimental parameters: v = 0.13 mm/h for
both cases, the imposed wavelength is 1.1 times the spontaneous
one in the case of the soliton solution case). }

\label{exp}

\end{figure}


\begin{thebibliography}{10}
\bibitem{footnote2}
Further analysis shows that this instability is related to the loss of a stable
fixed-point solution in the homogeneous phase diffusion equation (\ref{phase}).
\bibitem{footnote3}
We have also observed the transition from TW to MTW at 2:1 resonance.
\bibitem{footnote1}
The term 'soliton' is abused here in the sense of localized modulation, but the
referred states are not necessarily solitons in a strict sense.
\end{thebibliography}
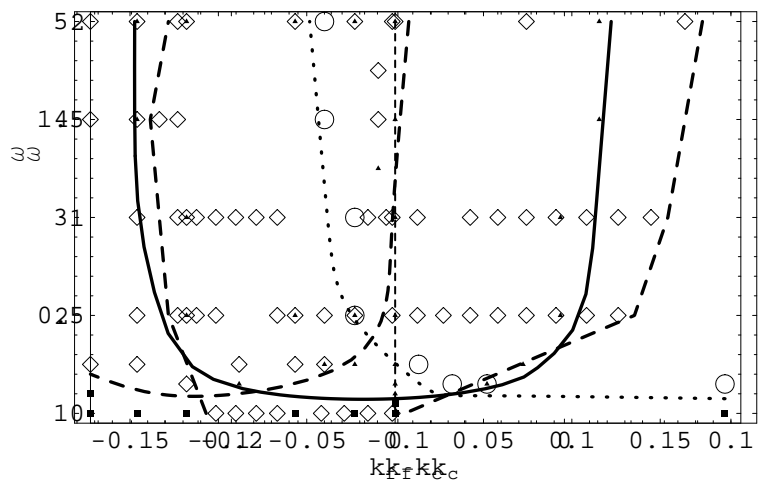
\end{document}